# AMRConvNet: AMR-Coded Speech Enhancement Using Convolutional Neural Networks

Williard Joshua Jose[12]
[1]Samsung R&D Institute Philippines, Taguig City, Metro Manila, Philippines
[2]University of the Philippines, Quezon City, Metro Manila, Philippines
williard.jose@eee.upd.edu.ph

*Abstract*—Speech is converted to digital signals using speech coding for efficient transmission. However, this often lowers the quality and bandwidth of speech. This paper explores the application of convolutional neural networks for Artificial Bandwidth Expansion (ABE) and speech enhancement on coded speech, particularly Adaptive Multi-Rate (AMR) used in 2G cellular phone calls. In this paper, we introduce AMRConvNet: a convolutional neural network that performs ABE and speech enhancement on speech encoded with AMR. The model operates directly on the time-domain for both input and output speech but optimizes using combined time-domain reconstruction loss and frequency-domain perceptual loss. AMRConvNet resulted in an average improvement of 0.425 Mean Opinion Score – Listening Quality Objective (MOS-LQO) points for AMR bitrate of 4.75k, and 0.073 MOS-LQO points for AMR bitrate of 12.2k. AMRConvNet also showed robustness in AMR bitrate inputs. Finally, an ablation test showed that our combined time-domain and frequency-domain loss leads to slightly higher MOS-LQO and faster training convergence than using either loss alone.

## I. Introduction

For more than a century, humans have been exchanging speech through electrical signals starting with the telephone. While early calls used analog signals, most calls nowadays use digital signals, converted to/from speech signals using speech coding. There are two broad quality and bandwidth classifications of speech coders: narrowband coders and wideband coders. Narrowband coders assume the bandwidth of intelligible speech is only up to around 4 kHz. The low bandwidth results to low bitrates and reduced speech quality. Wideband coders on the other hand include higher frequency components of up to 16-20 kHz. Wideband coders capture more realistic-sounding speech, but also increase bitrates.

Speech coders can also be classified based on how they model speech: waveform coders and vocoders [1]. Waveform coders expect general audio waveforms as input, and thus can encode and decode arbitrary waveforms. In contrast, vocoders assume that the input waveform is speech, and hence rely on speech models. This prior information enables vocoders to have lower bitrates than waveform coders, but also distorts the coded speech and makes them sound more artificial.

G.711 is a waveform coder which uses companded pulse code modulation (PCM). This enables it to accommodate a larger dynamic range of sample values [2]. It is slightly more compressed compared to regular PCM, although it still suffers from distortions such as quantization noise.

On the other hand, Adaptive Multi-Rate (AMR) is a vocoder developed by the 3GPP for efficiently encoding speech in phone calls while maintaining intelligibility [3]. AMR is currently being used in 2G phone calls. To support robustness on unpredictable phone signal strength, AMR transmits at variable bitrates: 4.75, 5.15, 5.90, 6.70, 7.40, 7.95, 10.20, and 12.20 kbit/s. The lower the bitrate, the lower the speech quality. In addition to having quantization noise, AMR uses a speech model that also distorts speech in more complex ways compared to waveform coders like G.711.

In general, speech encoded by compressive and lossy speech coders will lose some information and quality [1]. This is especially evident in older systems using narrowband speech coders, like the 30-year-old 2G cellphone system using AMR. Newer standards from 3G, 4G, and 5G are being developed which support wideband speech coders to support higher quality speech. But even as operators continue deploying new equipment with the latest standards, voice calls still fall back to narrowband speech coders if any device along the communication path does not support wideband speech coders.

The objective of this paper is to find ways of improving voice call quality encoded using AMR. In this paper, we design and train a convolutional neural network model which performs artificial bandwidth expansion and speech enhancement on AMR-coded speech.

Here are the contributions of this paper:

- To the best of our knowledge, this is the first work which uses convolutional neural networks to enhance speech encoded with AMR. This is a more challenging task compared to previous work on enhancing G.711-coded speech [4].

- We show how using a model trained with speech coded with higher bitrates can still lead to dramatic speech quality improvement when used to enhance speech coded with other bitrates.

- We note and compare the complementary relationship of time-domain reconstruction loss and frequency-domain perceptual loss. We illustrate how jointly using both losses increase model accuracy and achieve faster model training convergence.

## II. RELATED WORK

### A. Convolutional Neural Networks (CNN) for Image Super-Resolution

In recent years, deep learning has been widely successful in many computer vision tasks. One of these tasks is image super-resolution, where a 2D low resolution image must be upscaled to a higher resolution with no additional inputs provided [5]. End-to-end convolutional neural networks have been shown to have significant upscaling capability while being simple to implement and run [5]–[7]. Others used generative adversarial networks (GANs) [8]–[10] which perform photo-realistic super-resolution. These works can increase the image resolution from 1x scale up to 4x-16x scale. Similar techniques have been used to upscale videos [11], [12].

Signal processing techniques have become a standard tool across different domains such as electronics, image and video, and audio. Similarly, the success of deep learning for super resolution in the visual domain can be applied to other kinds of signals, in this case specifically for enhancement of coded speech signals.

### B. Artificial Bandwidth Expansion and Speech Enhancement

Artificial bandwidth expansion (ABE) is the problem of expanding bandwidth of speech from narrowband to wideband (20 kHz or more). Classical methods to solve this problem operate in the frequency-domain, by extending the magnitude and phase spectra of the audio signal [13], [14]. These methods have been improved by applying neural networks to do the bandwidth expansion in the frequency-domain [15]–[17].

There have been several recent works on artificial bandwidth expansion on time-domain signals, optionally with input from frequency-domain features. These use different architectures: regular convolutional neural networks [18] (from which this work is inspired from), time-frequency networks (TFNet) [19], and architectures inspired by fast Fourier transform algorithms (FFTNet) [20]. Performance in artificial bandwidth expansion has been steadily improving. However, these techniques generate the low resolution either by direct down sampling or by applying classical low pass filters.

The same techniques may not be directly applicable with coded speech, since speech coders also result in quantization noise, distortions, and coding artifacts. One recent work explored applying convolutional neural networks for the G.711 speech coder where they achieved state-of-the-art results [4]. However, G.711 is one of the simplest speech coders, being a waveform coder. Applying artificial bandwidth expansion and other speech enhancement techniques to more complex vocoders like AMR is more challenging. In this work, we trained a model to perform artificial bandwidth expansion and speech enhancement for the AMR speech coder.

## III. METHODOLOGY

### A. Dataset

The VCTK dataset [21] is a collection of speech utterances from native English speakers. This dataset contains 44 hours of speech utterances from 109 speakers. The speakers read out approximately 300-400 English sentences each.

High-quality speech samples were obtained from the VCTK dataset. These high-quality speech utterances originally have a sampling rate of 48 kHz. We down sampled this to 16 kHz .wav PCM files and used this as the reference ground truth speech.

To generate the lower-quality and bandwidth-limited coded speech, we used FFMPEG and OpenCORE AMR. The ground truth speech is encoded into AMR by FFMPEG, which is then decoded back to linearly coded .wav PCM format. Due to AMR's band limiting, the output sampling rate is only 8 kHz.

In our experiments, we selected 8 speakers to use for training. 80% of their speech samples were added to the training set, 10% were added to the validation set, while the remaining 10% were added to the test set. We ensured that the model only tests on previously unseen speech samples.

### B. Model Architecture

We posed this problem as a multivariate regression task: given $X$ coded speech samples and $Y$ high-quality ground truth speech samples, we would like to learn a transformation function $p_\theta(X)$ that outputs $\hat{Y}$ predicted speech samples. $p_\theta$ is parameterized by non-linear weights and biases $\theta$:

$$\hat{Y} = p_\theta(X) \qquad (1)$$

The transformation function $p_\theta$ can be modeled by a neural network of sufficient capacity. We adapted the U-Net model [22] and modified it to support 1-dimensional input and output signals [18]. The U-Net is an autoencoder architecture having an encoder $g_\theta$ and decoder $f_\theta$:

$$Z = g_\theta(X) \qquad (2)$$

$$\hat{Y} = f_\theta(Z, X) \qquad (3)$$

$$\hat{Y} = p_\theta(X) = f_\theta(g_\theta(X), X) \qquad (4)$$

The encoder compresses the coded speech samples $X$ into a latent vector $Z$. $g_\theta(X)$ can implicitly learn important features about $X$ such as phonemes, syllables, pitch, and timbre, and encode this into $Z$.

The decoder takes the latent code and generates speech with a higher quality and sampling rate. The U-Net architecture includes skip connections for each corresponding level of hierarchy in the encoder and decoder. This reframes the problem of generating enhanced speech samples from the latent vector, into the problem of *modifying* coded speech samples into enhanced speech samples. The skip connections also allow the model to transfer features common to both the low-quality and high-quality speech. Hence, $g_\theta(Z, X)$ also takes the coded speech samples $X$ as an input in addition to the latent vector $Z$.

We call our overall neural network architecture AMRConvNet (Fig. 1). Here, the encoder is implemented with six 1-dimensional convolutional layers (CONV), with the number of kernels per layer [256, 512, 512 512, 512, 512] and kernel sizes [65, 33, 33, 17, 9, 9]. Each CONV layer is followed by a dropout layer (DROP), which is important for generalizing especially for speech signals so that the model does not overfit with the training set's phase. Each layer is also zero-padded and has a stride of 2 to halve the number of outputs with increasing layer count. The output of the last encoder CONV/DROP layers is the latent vector, which is also the input to the decoder. The decoder is implemented following the encoder (same number of layers, kernel sizes, and kernel counts), but in reverse, and each succeeding layer is sub-sampled so that it has twice the number of outputs with increasing layer count.

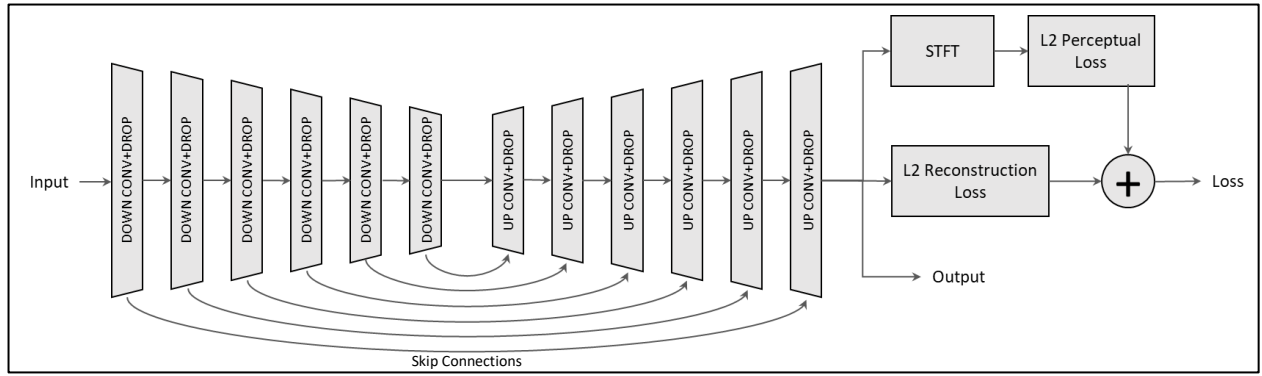

Fig. 1. AMRConvNet neural network architecture.

*C. Loss Function*

We used a combination of time-domain L2 loss and frequency-domain L2 loss for encouraging AMRConvNet to converge. Mean-squared error (MSE) on the time-domain signals encourage AMRConvNet to regress the original higher frequency and undistorted time-domain signal. This time-domain MSE is termed reconstruction loss, as we are trying to reconstruct the original time-domain signals.

The frequency-domain MSE is obtained by getting the short-time Fourier transform (STFT) magnitude of both the original speech signal and the predicted speech signal resulting in two 2-dimensional arrays whose element-wise square error is averaged. The result, herein termed as perceptual loss, is added linearly to the time-domain (Fig. 1). The combined loss is then the sum of the reconstruction loss and perceptual loss:

$$\mathcal{L} = \frac{1}{T}\sum_{i=1}^{T}(\hat{y}_i - y_i)^2 + \lambda \times \frac{1}{TN}\sum_{i=1}^{T}\sum_{j=1}^{N}(\hat{h}_{ij} - h_{ij})^2 \quad (5)$$

where $T$ denotes the number of samples, $N$ denotes the number of frequency points, $\hat{y}_i$ denotes output time-domain signals, $y_i$ denotes ground truth time-domain signals, $\hat{h}_{ij}$ denotes output magnitude spectrum values, and $h_{ij}$ denotes ground truth magnitude spectrum values. Weighting of importance between reconstruction and perceptual loss is controlled by the hyper-parameter scalar $\lambda$.

One possible problem with purely time-domain reconstruction is phase variation. Two signals with a phase offset can have significantly different samples (e.g. when offset is 90°) but are perceptually similar. This can be alleviated by adding a frequency-domain MSE, since the magnitude spectrum is not affected by phase variation. MSE on the frequency-domain signals also allows AMRConvNet to explicitly recover high frequency components of the original speech signal which were discarded when encoding was applied. At the same time, it corrects low-frequency distortions introduced by the speech coder, with the solution not needing to have the same phase spectrum as the original speech signal.

*D. Training and Validation*

Optimization of the neural network was performed using the Adam optimizer, with learning rate set to 3e-4. Training was performed to run up to 300 epochs with early stopping.

We also performed a hyperparameter search on: (1) the number of convolutional layers in the encoder and decoder; (2) the kernel size of each 1D convolutional layer; (3) the number of kernels per convolutional layer; and (4) the hyperparameter $\lambda$ weighting the two loss functions. This search led to the model architecture previously described.

We implemented the neural network in Keras and TensorFlow. We trained AMRConvNet using a workstation with i7-7700k, 32GB RAM, and 1x RTX 2080Ti GPU. Each model iteration training ran for around 22 hours.

*E. Speech Quality Testing Metrics*

Speech quality can be compared subjectively and objectively. The most common subjective test is Mean Opinion Score (MOS). MOS is a qualitative score from a scale of 1-5 which describe how good or how bad the quality of speech is (1 being Bad and 5 being Excellent) [23]. The official ITU-T recommendation prescribes having multiple volunteers listen and score the speech samples from 1-5 in a room of size 30-120 cubic meters. Having an MOS of 4.0 or better is usually considered high quality [24].

Speech quality can also be measured objectively with purpose-built algorithms. The most widely used is the ITU-T metric: Perceptual Evaluation of Speech Quality (PESQ) [25]. PESQ is a family of objective tests which perform mathematical transformations on a speech signal to derive a numeric value from -0.5 to 4.5. PESQ also prescribes a mapping function from the normal PESQ metric to its equivalent MOS-LQO (mean opinion score – listening quality objective) value from 1.02 to 4.56. For most applications in speech, PESQ is a reliable objective metric of speech quality [26] and intelligibility [27], and has been found to correlate well with MOS.

To compare the speech quality of two speech samples, we used PESQ. PESQ was chosen over MOS to maintain stable and repeatable speech quality results across all speech samples, as well as to allow rapid iteration and more tests without needing to wait for MOS results. PESQ results are also comparable among different works, whereas MOS cannot be directly compared among different works since most works use different sets of evaluators as well.

PESQ evaluates the score between two speech files: one is the ground truth (high-quality) speech, and the other is the degraded (low-quality) speech. We then find the mean PESQ MOS-LQO score across all speech utterances for a given speaker or set of speakers to compute the final score.

## IV. RESULTS

We trained AMRConvNet and evaluated its performance with multiple speakers and on the multiple bitrates supported by AMR. Then we investigated how AMRConvNet generalizes with varying bitrates. This was done to check the robustness of AMRConvNet, given that 2G phone calls in the real world can dynamically change bitrates depending on phone signal quality. Finally, we performed an ablation test between the reconstruction loss (time-domain) and the perceptual loss (frequency-domain).

### A. Evaluation over Multiple Speakers

We trained and measured the performance of the model on eight speakers at two bitrates (4.75k and 12.2k). These two bitrates were chosen because these are the minimum and maximum bitrates supported by AMR. This was done to investigate how model performance is affected by the speech quality of the input itself. We compare the PESQ MOS-LQO scores of the original bandlimited AMR speech samples with those of the enhanced speech samples (Table 1). Across all speakers and bitrates, the average MOS-LQO scores increased.

For bitrate 4.75k, the average improvement is 0.425 MOS-LQO points, which is a significant 13% increase. This makes sense as the speech coder throws away more information on the speech characteristics as the bitrate decreases. For the maximum bitrate 12.2k on the other hand, the average improvement is lower at 0.073 MOS-LQO points (1.79% increase). This is because 12.2k bitrate already captures the quality of the speech well (at an average of 4.069 MOS-LQO points for the coded speech input). The higher frequencies that were reconstructed by the model might not have a significant effect on the perception of quality.

There is some variability in the model's performance with different speakers (Fig. 2-3). This is partly caused by differences in the quality of voice of some speakers. Other voices might have more complex features which are harder to model and generalize on.

We also checked the quality of the enhanced speech by inspecting their magnitude spectrograms generated using short-time Fourier transform (STFT). Most samples show an extension of the frequency content of the speech waveforms beyond the original AMR coded speech (Fig. 4). We can infer that the model was able to recognize patterns from the low-frequency coded speech and mapped them to the higher-frequency speech, since the waveforms look similar in structure to the ground truth speech.

TABLE I. AVERAGE MULTI-SPEAKER PESQ MOS-LQO SCORES

| AMR Bitrate | PESQ MOS-LQO | | |
|---|---|---|---|
| | AMR-Coded Input Speech | Enhanced with AMRConvNet | Improvement |
| 4.75k | 3.175 | 3.583 | 0.425 |
| 12.2k | 4.069 | 4.142 | 0.073 |

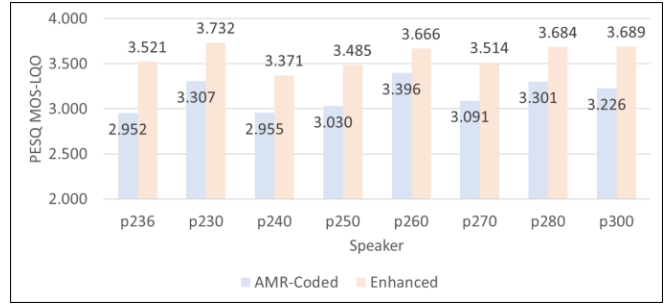

Fig. 2. Multi-speaker performance for AMR bitrate 4.75k.

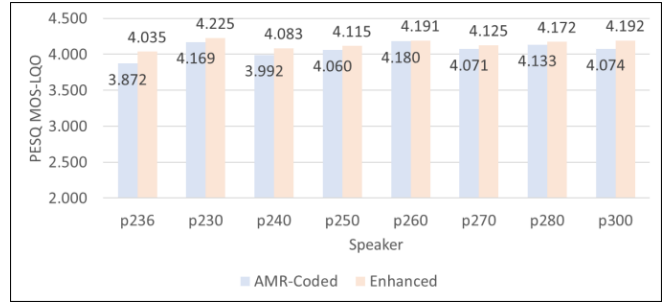

Fig. 3. Multi-speaker performance for AMR bitrate 12.2k.

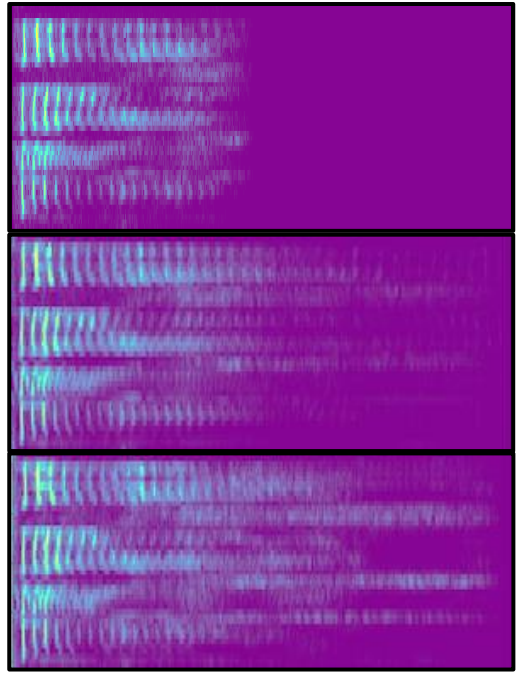

Fig. 4. Sample speech spectrogram. Top: band-limited AMR speech signal. Middle: enhanced and up sampled speech signal. Bottom: ground truth reference speech signal.

## B. Evaluation over AMR Bitrates

We investigated the effect of varying input bitrates to a model pre-trained on a specific bitrate. For this experiment, we trained the model on speech encoded with 12.2k AMR. We tested the model on speech from two speakers, p236 and p300, whose speech samples were encoded in 4.75k, 5.15k, 5.9k, 6.7k, 7.4k, 7.95k, 10.2k, 12.2k (Table 2-3).

We can see from the trend that output speech quality is monotonically increasing with increasing input bitrates (Fig. 5-6). This shows that the model is adaptive and robust to changes in the bitrate input and maximizes the enhancement it can perform from the given input.

## C. Ablation Test: Reconstruction and Perceptual Loss

To investigate the individual contributions of the reconstruction (time-domain) and perceptual (frequency-domain) losses, we trained the model with each loss alone. We performed this test with input AMR bitrate of 4.75k.

Our results show that the combined loss has slightly better PESQ MOS-LQO scores (Table 4). This suggests that both time-domain and frequency-domain losses allow the model to approach its performance limit.

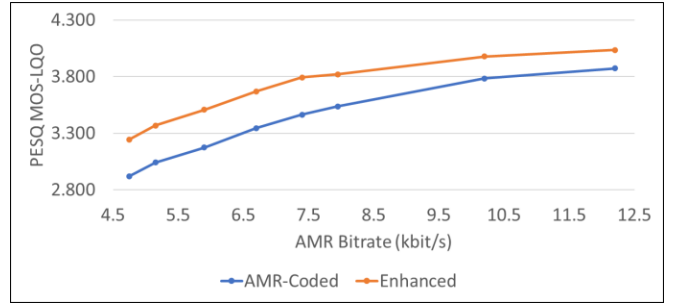

Fig. 5. Effect of input bitrate on model performance for speaker p236.

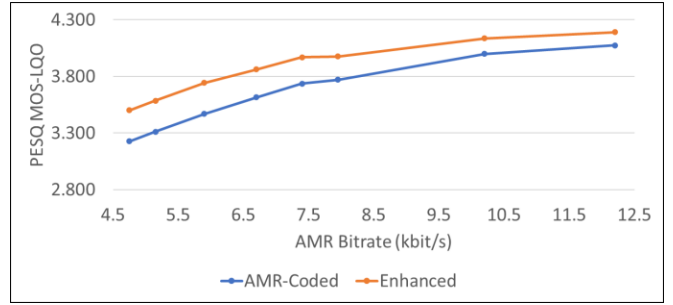

Fig. 6. Effect of input bitrate on model performance for speaker p300.

TABLE II. PESQ MOS-LQO SCORES FOR SPEAKER P236 WITH VARYING INPUT BITRATES

| Bitrate | PESQ MOS-LQO | | |
|---|---|---|---|
| | *AMR-Coded Input Speech* | *Enhanced with AMRConvNet* | *Improvement* |
| 4.75k | 2.920 | 3.244 | 0.323 |
| 5.15k | 3.041 | 3.368 | 0.328 |
| 5.9k | 3.173 | 3.507 | 0.333 |
| 6.7k | 3.344 | 3.669 | 0.325 |
| 7.4k | 3.464 | 3.793 | 0.329 |
| 7.95k | 3.537 | 3.819 | 0.283 |
| 10.2k | 3.784 | 3.978 | 0.194 |
| 12.2k | 3.872 | 4.035 | 0.164 |

TABLE III. PESQ MOS-LQO SCORES FOR SPEAKER P300 WITH VARYING INPUT BITRATES

| Bitrate | PESQ MOS-LQO | | |
|---|---|---|---|
| | *AMR-Coded Input Speech* | *Enhanced with AMRConvNet* | *Improvement* |
| 4.75k | 3.226 | 3.502 | 0.276 |
| 5.15k | 3.312 | 3.586 | 0.274 |
| 5.9k | 3.469 | 3.743 | 0.274 |
| 6.7k | 3.616 | 3.862 | 0.246 |
| 7.4k | 3.736 | 3.970 | 0.234 |
| 7.95k | 3.771 | 3.976 | 0.205 |
| 10.2k | 4.000 | 4.135 | 0.135 |
| 12.2k | 4.074 | 4.192 | 0.118 |

TABLE IV. ABLATION TEST PESQ MOS-LQO SCORES FOR RECONSTRUCTION LOSS, PERCEPTUAL LOSS, AND COMBINED LOSS

| Loss | PESQ MOS-LQO | | |
|---|---|---|---|
| | *AMR-Coded Input Speech* | *Enhanced with AMRConvNet* | *Improvement* |
| Reconstr. | 3.226 | 3.668 | 0.441 |
| Percept. | 3.226 | 3.668 | 0.441 |
| **Combin.** | **3.226** | **3.689** | **0.462** |

In addition, looking at the loss trajectories, we can see that the model converges much faster (in terms of reconstruction loss, perceptual loss, and PESQ MOS-LQO trajectories) when training with the combined loss vs either loss alone (Fig. 7-9). This suggests better training trajectories due to the losses acting as regularizers to each other.

## V. CONCLUSION

In this paper, we proposed a convolutional neural network architecture operating directly on the time-domain and optimized by combined reconstruction and perceptual losses. The model enhances narrowband speech encoded using Adaptive Multi-Rate (AMR) speech coder. We demonstrated the ability of AMRConvNet to increase the quality of coded AMR speech across different speakers. We also showed the robustness of AMRConvNet to varying input speech coder bitrates. Finally, we showed how our combined time-domain and frequency-domain loss improved the output speech quality and led to significantly faster training times. AMRConvNet resulted in an average improvement of 0.425 MOS-LQO points for AMR bitrate of 4.75k and 0.073 MOS-LQO points for AMR bitrate of 12.2k. Future work can experiment with more complex models and propose new architectures to better handle the time-domain and frequency-domain duality of audio signals.

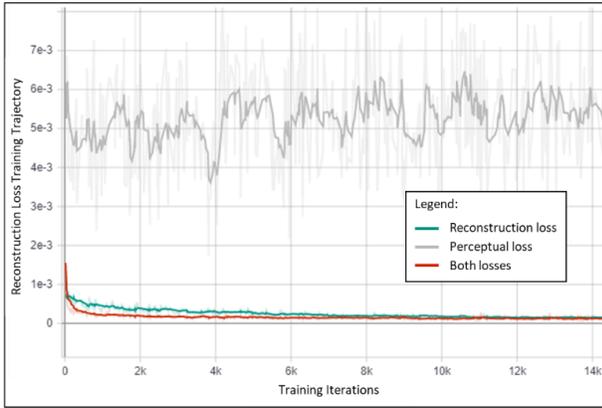

Fig. 7. Comparison of reconstruction loss trajectory for ablation test.

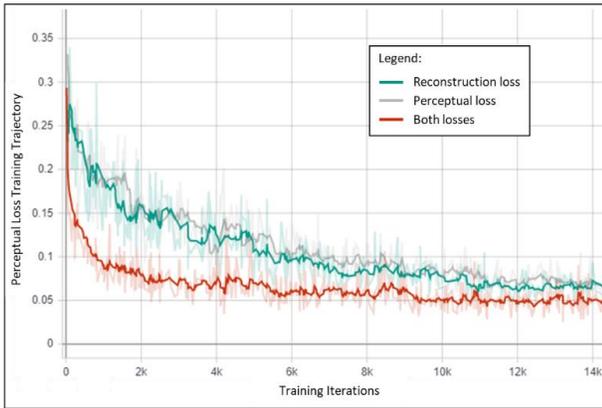

Fig. 8. Comparison of perceptual loss trajectory for ablation test.

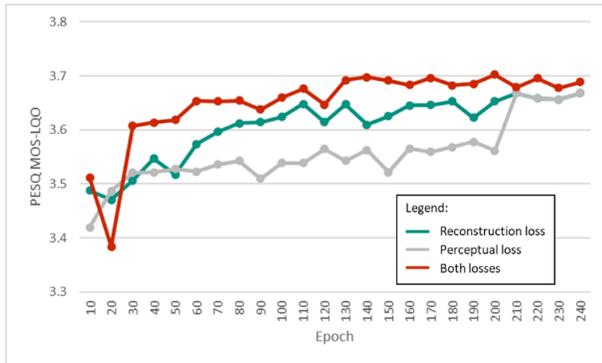

Fig. 9. Comparison of PESQ MOS-LQO trajectory for ablation test.